\pdfoutput=1
\documentclass[twocolumn,english,aps,superscriptaddress,prb]{revtex4-1}

\usepackage{babel}
\usepackage{amsmath}
\usepackage{amssymb}
\usepackage{graphicx}

\newcommand{\be}{\begin{equation}}
\newcommand{\ee}{\end{equation}}
\newcommand{\bea}{\begin{eqnarray}}
\newcommand{\eea}{\end{eqnarray}}

\begin{document}

\title{Dimerization transitions in spin-1 chains}

\author{Natalia Chepiga}
\affiliation{Institute of Physics, Ecole Polytechnique F\'ed\'erale de Lausanne (EPFL), CH-1015 Lausanne, Switzerland}
\author{Ian Affleck}
\affiliation{Department of Physics and Astronomy, University of British Columbia, Vancouver, BC, Canada V6T 1Z1}
\author{Fr\'ed\'eric Mila}
\affiliation{Institute of Physics, Ecole Polytechnique F\'ed\'erale de Lausanne (EPFL), CH-1015 Lausanne, Switzerland}

\date{\today}
\begin{abstract} 
We study spontaneous dimerization transitions in a Heisenberg spin-1 chain with additional next-nearest neighbor (NNN) and 3-site 
interactions using extensive numerical simulations and a conformal field theory analysis. We show that the transition
can be second order in the WZW SU(2)$_2$ or Ising universality class, or first-order.
We argue that these features are generic because of a marginal operator in the WZW SU(2)$_2$ model, and because
of two topologically distinct non-dimerized phases with or without edge states. We also provide explicit numerical evidence 
of conformal towers of singlets inside the spin gap at the Ising transition. Implications for other models  are briefly discussed.
\end{abstract}

\maketitle


Topological matter is currently attracting a lot of attention. One of the first examples is the spin-1 Heisenberg
chain, which has a bulk gap\cite{haldane} but spin-1/2 edge states.\cite{kennedy,hagiwara} Spin-1 chains with more
general interactions have been extensively studied over the years, and they have in particular been shown 
to undergo a spontaneous dimerization in the presence of a negative biquadratic interaction at an integrable
critical point.\cite{babujian,takhtajan} The universality class of this critical point is SU(2)$_2$ 
Wess-Zumino-Witten (WZW) with central charge $c=3/2$.\cite{affleck86_1,affleck86_2,affleck_haldane} It has been identified 
in other models exhibiting spontaneous dimerization,\cite{michaud1} and it is usually considered to describe the 
generic behaviour of spin-1 chains at the transition to a spontaneously dimerized phases.

In this Letter, we identify two other generic possibilities, Ising and first order, and we show that these alternatives
are natural consequences of general properties: i) the presence of topological and
non-topological phases with and without edge states respectively; ii) the existence of a marginal operator in the 
WZW SU(2)$_2$ model. We also show that combining density matrix renormalization group (DMRG) simulations
with conformal field theory (CFT) predictions for {\it open} systems gives access to the conformal towers of the critical
lines, including that of singlets inside the spin gap on the Ising line.

We  consider the spin-1 chain Hamiltonian:
\begin{multline}
\label{eq:H}
H=\sum_{i}\left(J_1{\bf S}_i\cdot {\bf S}_{i+1}+J_2{\bf S}_{i-1} \cdot {\bf S}_{i+1}\right)\\
+\sum_{i}J_3\left[({\bf S}_{i-1}\cdot {\bf S}_i)({\bf S}_i\cdot {\bf S}_{i+1})+{\mathrm H.c.}\right]
\end{multline}
On top of the standard Heisenberg coupling $J_1$, it includes two of the three interactions that appear in next-to-leading order in the strong coupling expansion of the two-band Hubbard model: the NNN interaction $J_2$ and a three-site interaction with coupling strength $J_3$. (The biquadratic interaction $({\bf S}_i\cdot {\bf S}_{i+1})^2$ has been omitted for simplicity.)
We set $J_1=1$ throughout the paper and concentrate on the case $J_2,J_3\geq0$.

\begin{figure}[h!]
\includegraphics[width=0.47\textwidth]{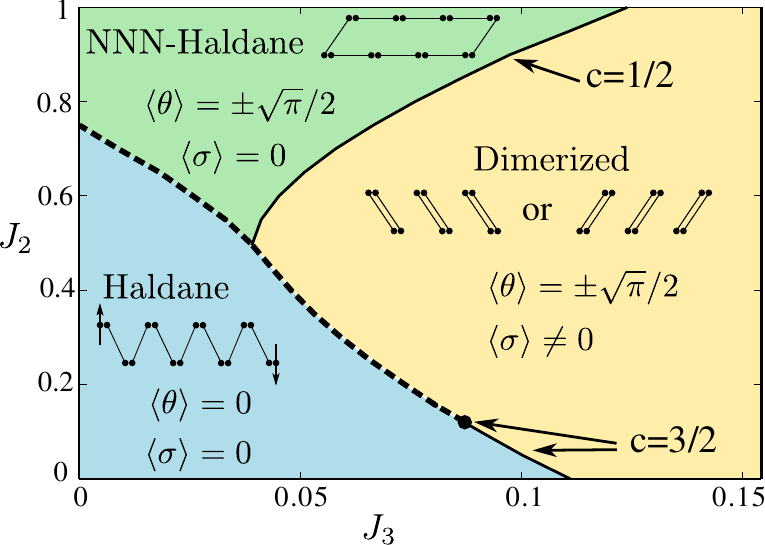}
\caption{(Color online) Phase diagram of the spin-1 chain with next-nearest neighbor coupling $J_2$ and 3-site interaction $J_3$. The transition from the dimerized phase to the Haldane phase is continuous along the solid line, with central charge $c=3/2$, and first order along the dashed line. The transition from the NNN-Haldane to the dimerized is a continuous transition in the Ising universality class with central charge $c=1/2$. The transition between the Haldane phase and the NNN-Haldane phase is always first order.}
\label{fig:phase_diagram}
\end{figure}

Let us first summarize the main results obtained using extensive DMRG simulations and
exact diagonalizations (ED). The phase diagram as a function of $J_2$ and $J_3$ consists of three phases, each of which may be schematically illustrated 
by a  diagram with lines indicating valence bond singlets formed between various site, (see Fig.~\ref{fig:phase_diagram}): a Haldane phase with one valence bond 
per $J_1$ bond, a next-nearest neighbor (NNN)-Haldane phase with one valence-bond  per $J_2$ bond, and a dimerized phase with two
valence-bonds on every other $J_1$ bond. The characterization of the short-range correlations (including disorder and Lifshitz lines) will be reported elsewhere.\cite{chepiga}

The transition between the Haldane and the NNN Haldane phase is always first order (the energy per site has a kink), in agreement with previous results for $J_3=0$.\cite{kolezhuk_prl}  It is topological: the two phases cannot be distinguished by any local order parameter, but the Haldane phase is topological (supports gapless edge states), while the NNN-Haldane is not
(see sketches in Fig.~\ref{fig:phase_diagram}). 

For small $J_2$, the transition between the Haldane and Dimerized phases is in the SU(2)$_2$ WZW universality class with central charge
$c=3/2$ from $J_2=0,J_3\simeq 0.111$\cite{michaud1} up to (and including at) a critical end point beyond which the transition becomes first order 
(see below).
There is actually a simple argument in favor of a first order transition 
in this parameter range: The fully dimerized state is an exact ground state at $J_2=0,J_3=1/6$,\cite{michaud1} and it remains an exact eigenstate
along the line $J_2+3J_3=1/2$,\cite{wang} but it is {\it not} the ground state at $J_2=1/2,J_3=0$. So a first-order transition
where the dimerization disappears abruptly has to take place. 
This first-order line connects smoothly, at an unusual triple point, with 
the first-order transition between the Haldane and NNN Haldane phases.\cite{chepiga}

\begin{figure}[h!]
\includegraphics[width=0.47\textwidth]{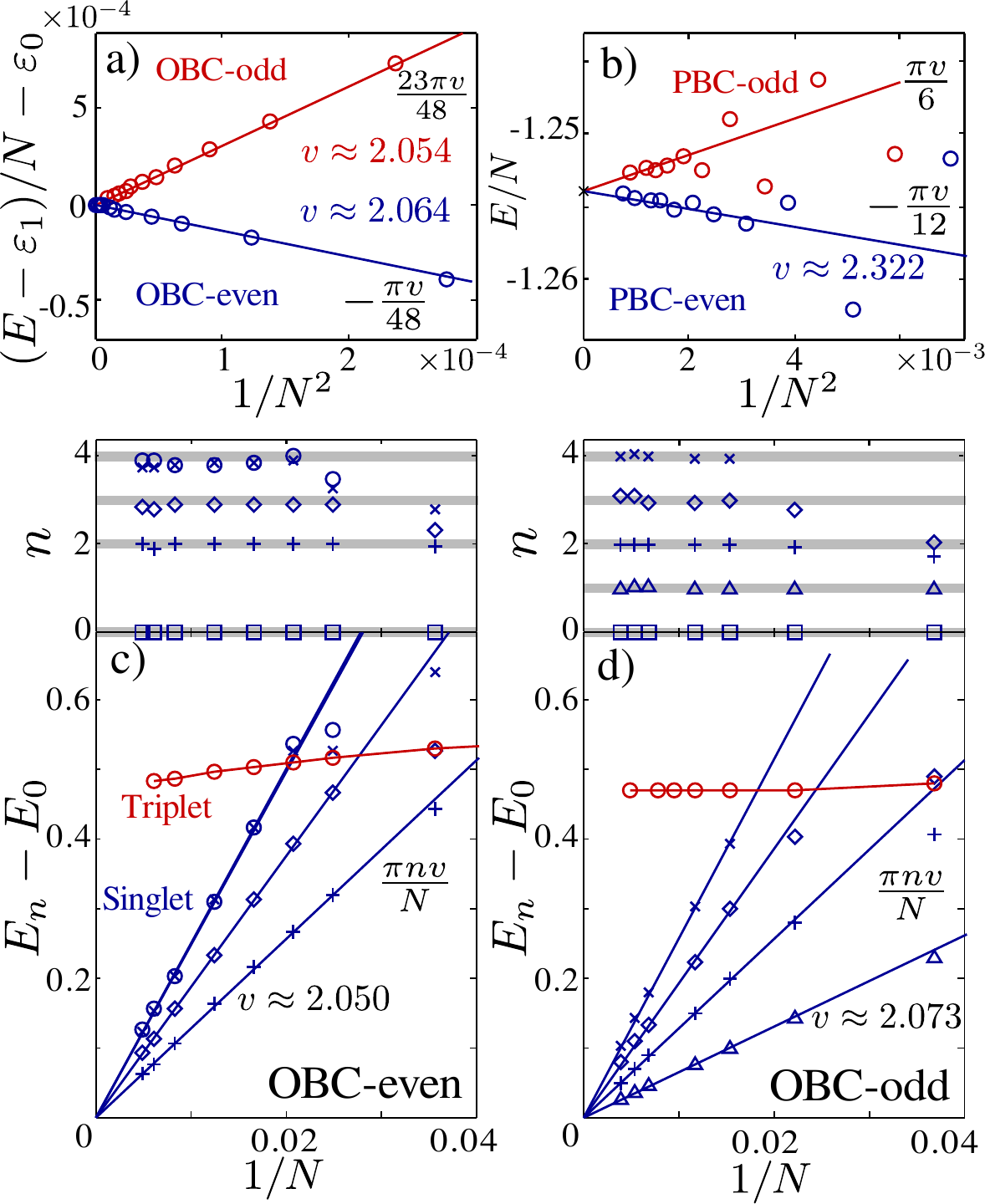}
\caption{(Color online) Ground state and excitation energy at $J_2=0.7$ and $J_3=0.058$, on the Ising line. 
 a) Linear scaling of the ground state energy per site in open chain with $1/N^2$ after subtracting $\varepsilon_0$ and $\varepsilon_1$ terms. b) Linear scaling of the 
 ground state energy per site with $1/N^2$ for periodic chain. 
c) and d) Energy gaps in singlet and triplet sectors for OBC as a function of $1/N$ for even and odd number of sites. The slope of singlet gap gives values of the velocity. Inset: Conformal towers. Grey lines show Ising conformal towers $I$ ($N$ even) and $\epsilon$ ($N$ odd); blue symbols are DMRG data.}
\label{fig:energy}
\end{figure}

Finally, the transition between the NNN Haldane and Dimerized phases is in the Ising universality class. 
As shown in Fig.~\ref{fig:energy}, singlet excitations become critical and build a conformal tower while the magnetic excitations remain gapped at the transition.
Note that we have reached similar conclusions regarding the phase transitions for the spin-1 chain with NNN and
biquadratic interactions,\cite{chepiga} in partial disagreement with Pixley et al \cite{nevidomskyy}, who in particular reached the conclusion that the transition between the NNN Haldane
and Dimerized phase is first order.

This phase diagram and the nature of the various transitions can be understood using conformal field theory (CFT) techniques.  We begin near the $SU(2)_2$ critical point 
where the low energy degrees of freedom of the spin chain can be written in terms of an $SU(2)$ matrix field, $g(x,t)$. The staggered component of the spin 
operators become $\vec S_j\propto (-1)^j\hbox{tr}\vec \sigma g(j)$ and the dimerisation operator becomes $\vec S_j\cdot \vec S_{j+1}\propto (-1)^j\hbox{tr}g(j)$. 
The low energy effective Hamiltonian  is that of the $SU(2)_2$ WZW model together with one relevant and one marginal operator, 
\be {\cal H}={\cal H}_{WZW}+\lambda_1 (\hbox{tr} g)^2+\lambda_2 \vec J_R\cdot \vec J_L,\ee
 where $\vec J_{L/R}$ are the left (right) moving spin currents. 
The relevant coupling constant, $\lambda_1$, controls the Haldane to Dimerized transition. When $\lambda_1<0$, $\langle \hbox{tr} g\rangle$ becomes non-zero corresponding to 
dimerisation.\cite{affleck_haldane}  When $\lambda_1>0$ $\langle \hbox{tr} g\rangle =0$ corresponding to the Haldane phase.  The marginal coupling constant, $\lambda_2$, 
renormalises to zero if it is initially negative.  In this regime the Haldane to Dimerized transition is second order, with the WZW model  occurring along the critical line with logarithmic corrections to scaling.  These logarithmic corrections vanish at the end of the critical line, where $\lambda_2=0$.\footnote{The end of the critical line does {\it not} correspond to a $k>2$ WZW model, but simply to the point where $k=2$ WZW behavior is most clearly 
observed,  due to the 
vanishing of the marginally irrelevant coupling constant there.  We disagree in this regard with Ref.\onlinecite{nevidomskyy}.} 
When $\lambda_2>0$ it renormalises to large values. 

To understand the full phase diagram, it is very useful to use a conformal 
embedding (also called a coset construction), an exact representation of the $SU(2)_2$ WZW model as a direct product of a free boson and an Ising model.\cite{fateev}
 All operators in the WZW model can 
be represented as products of free boson and Ising operators. In particular:\cite{supmat}
\bea 
&&\hbox{tr}g \propto \sigma \sin \sqrt{\pi} \theta ,\ \ 
(\hbox{tr}g)^2\propto \epsilon -C_1\cos \sqrt{4\pi}\theta \nonumber \\
&&\vec J_L\cdot \vec J_R\propto \epsilon \cos \sqrt{4\pi}\theta +C_2\partial_x\phi_L\partial_x\phi_R\label{CE}\eea
 for constants $C_1$ ($>0$) and $C_2$. 
To see how $\lambda_1$ induces the Haldane to Dimerized transition, note that a positive $\lambda_1$ pins $\theta$ at $0$ whereas a negative 
$\lambda_1$ pins it at $\pm \sqrt{\pi}/2$, leading to  $\langle \sin \sqrt{\pi}\theta \rangle \neq 0$.   At the same time, a positive coefficient of $\epsilon$ in the Ising Hamiltonian corresponds to the disordered phase 
whereas a negative coefficient to the ordered phase with $\langle \sigma \rangle \neq 0$.  Thus we obtain, from Eq. (\ref{CE}),
 $\langle \hbox{tr}g\rangle \neq 0$ for $\lambda_1<0$.  
Remarkably, in this 
representation of the WZW model, a second order transition occurs simultaneously in Ising and boson sectors.  The first order transition for $\lambda_2>0$ 
can be understood intuitively in this representation. A large positive $\lambda_2$ favours states with $\langle \epsilon \rangle \langle \cos \sqrt{4\pi}\theta \rangle <0$.  There are then 
two degenerate gapped states with $\langle \epsilon \rangle <0$, $\theta$ pinned at $0$ corresponding to the Haldane phase or $\langle \epsilon \rangle >0$, $\theta$ pinned at $\pm \sqrt{\pi}/2$ corresponding to the dimerised phase.  Turning on $\lambda_1$ splits the degeneracy, leading to a first order transition.

So, far we have focussed on the vicinity of the WZW critical point.  Let us now consider what may happen as we move far from it along the first order transition line. It is 
now no longer permissible to only consider the couplings which are relevant at the critical point so the Ising and boson transitions could occur at different 
places in the phase diagram.  For instance, a $\lambda_3\cos 3\sqrt{4\pi}\theta$ term would favour either $\langle \theta \rangle =0$ or $\langle \theta \rangle =\pm \sqrt{\pi}/2$ 
depending on its sign.  If $\lambda_3$ changed sign along a line in the phase diagram the transition could occur in the boson sector without occurring simultaneously 
in the Ising sector.  This phase with $\langle \theta \rangle =\pm \sqrt{\pi}/2$, $\langle \sigma \rangle =0$ corresponds to the NNN Haldane phase. 
This can be seen from the presence of gapless S=1/2 edge excitations when $\langle \theta \rangle =0$ but not when $\langle \theta \rangle =\pm \sqrt{\pi}/2$. 
An open boundary favors a dimer ending at the last site. 
Hence $\langle \hbox{tr}g(x)\rangle$ becomes non-zero near the boundary. Thus  $\langle \theta (x)\rangle$ takes the value $\pm \sqrt{\pi}/2$ at an open end\cite{supmat}
 and $\langle \sigma (x)\rangle$ becomes non-zero.  However, 
$\langle \theta (x) \rangle =0$ far from the boundary in the Haldane phase. This rotation of $\theta (x)$ corresponds to an induced magnetisation:
\be  \sum_jS^z_j=\int _0^\infty dx (d\theta /dx)/\sqrt{\pi}=\pm 1/2\ee 
near an open boundary at $x=0$ in the Haldane phase. By contrast, there is no induced magnetization  in the NNN Haldane phase since 
$\langle \theta (x)\rangle =\pm \sqrt{\pi}/2$ in the bulk, so it doesn't rotate at the boundary. So this phase has no gapless edge modes but also 
has no dimerisation since $\langle \sigma \rangle =0$.  Thus, we may identify it with the NNN Haldane phase.  
We now see that a third transition can also take place in which $\theta$ remains pinned at $\pm \sqrt{\pi}/2$ while the sign of the $\epsilon$ term in the Hamiltonian 
changes. This corresponds to an Ising transition from NNN Haldane to Dimerized phases.  The gap in the boson sector, at this transition, implies 
a gap for all magnetic excitations.\cite{supmat}

\begin{figure}[h!]
\includegraphics[width=0.47\textwidth]{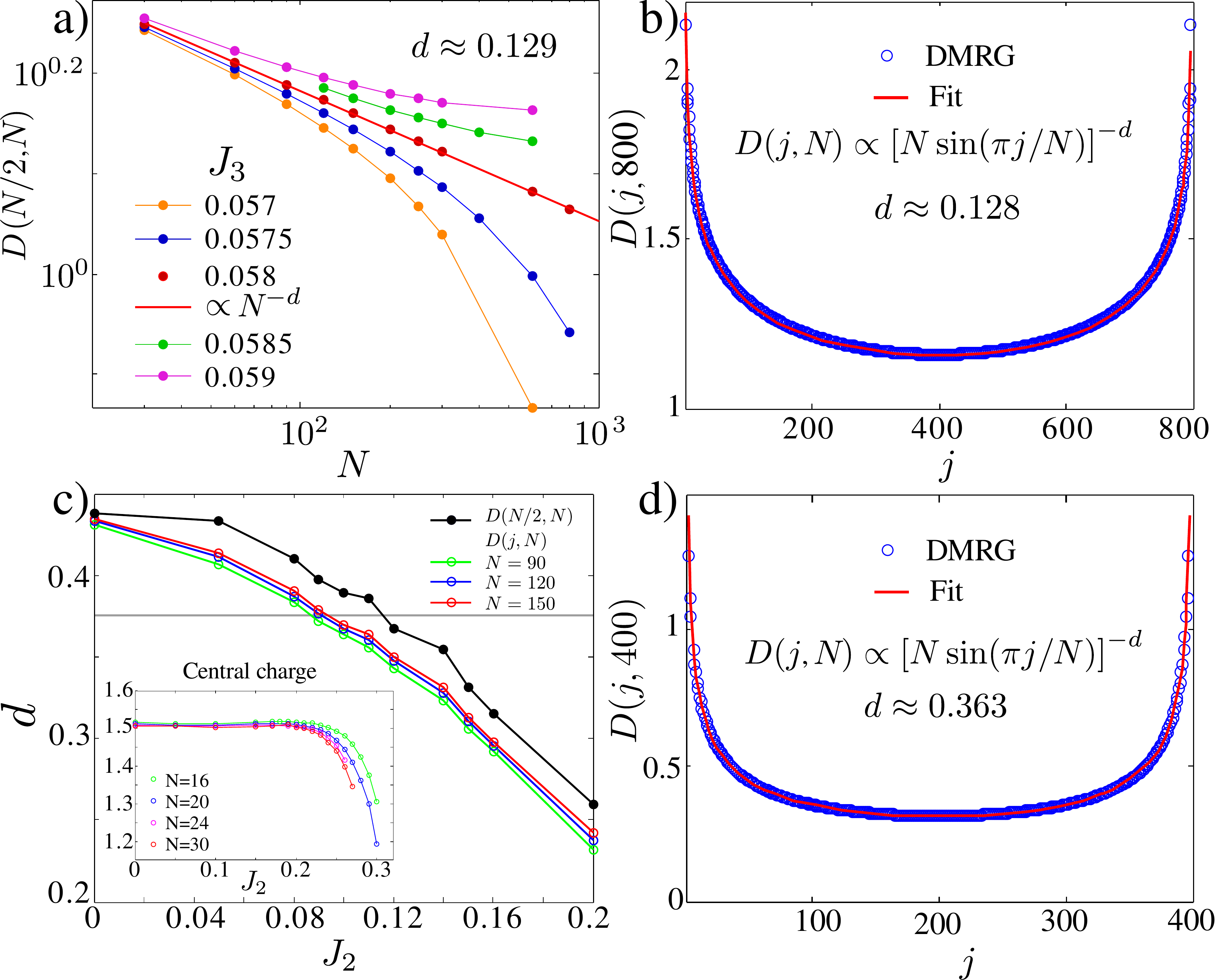}
\caption{(Color online) (a) Log-log plot of the mid-chain dimerization as a function of the number of sites $N$ for $J_2=0.7$ and different values of $J_3$. The 
linear curve corresponds to the Ising critical point, and the slope to the critical exponent. This leads to $J_{3c}=0.058$, and to a slope $0.129$, in good 
agreement with the prediction $1/8$ for Ising.
(b) Site dependence of $D(j,N)$ at the critical point fitted to $1/[N \sin (\pi j/N)]^{d}$.
This determines an exponent $d=0.128$, again close to the Ising prediction
$1/8$.  (c) Apparent critical exponent along the $SU(2)_2$ critical line as a function of $J_2$. Black solid circles: from the slope of the log-log plot $D(N/2,N)$
as a function of $N$ for the value of $J_3$ for which it is linear. Open color circles: from fitting $D(j,N)$ for different sizes at the same points. 
The dashed line is the theoretical value of the exponent, $3/8$. Inset: central charge along the critical line as determined from fitting the entanglement 
entropy of periodic chains with 
the Calabrese-Cardy formula. 
(d) $D(j,N)$ at the $SU(2)_2$ critical end point fitted to  $1/[N \sin (\pi j/N)]^{d}$. The exponent is in good agreement with $d=3/8$. 
}
\label{fig:dimerization}
\end{figure}

Let us now use CFT to extract more precise information about the phase diagram, beginning with the Ising transition. 
As discussed above, an  open boundary condition favors dimerisation, corresponding to a non-zero boundary magnetic field in the Ising model. It then follows from boundary CFT that the magnetization at the critical point decays away from the boundary 
as\cite{cardy91} $\langle \sigma (x)\rangle\propto 1/x^{1/8}$ since $1/8$ is the scaling dimension of $\sigma$. 
For a finite system of $N$ sites, a conformal transformation gives
$\langle \sigma (x)\rangle \propto 1/[(N/\pi) \sin (\pi x/N)]^{1/8}$.
 On a finite chain, we define the local dimerization as $D(j,N)=|\langle \vec S_j.\vec S_{j+1} \rangle - \langle \vec S_{j-1}.\vec S_j \rangle|$. 
Identifying the local dimerization with $\sigma$, this leads to  $D(j,N)\propto 1/[N \sin (\pi j/N)]^{1/8}$, and in particular to 
$D(N/2,N) \propto 1/N^{1/8}$.  Plotting $D(N/2,N)$ versus $N$ on a log-log plot, we determine the Ising critical line by the points where this 
curve is linear.  We find the slope is close to $1/8$ along the entire Ising critical line. An example of 
data on a line crossing the Ising critical line is shown in  Fig. \ref{fig:dimerization} (a). Along the  critical line we also find a good 
fit of $D(j,N)$ to $1/[N \sin (\pi j/N)]^{1/8}$ as shown in Fig. \ref{fig:dimerization} (b).

CFT predicts that all excitation energies, for any conformally invariant boundary conditions, are of the 
form $(\pi v/N)x_n$ where $v$ is the velocity and the dimensionless numbers $x_n$ are universal scaling dimensions of operators.\cite{cardy86}
Furthermore, the ground state energy contains 
a universal term $-\pi vc/(24N)$ for open boundary conditions (OBC) and $-\pi vc/(6N)$ for periodic boundary conditions (PBC) where $c$ is the central charge.   
We identify  OBC with $\uparrow$,$\uparrow$ boundary conditions in the Ising model for $N$ even and $\uparrow$,$\downarrow$ boundary conditions for 
$N$ odd, where the arrows refer to the directions of the boundary magnetic fields.  This follows because OBC favour the same sign of the dimerisation  
at both ends of the system for $N$ even but opposite signs for $N$ odd. Similarly we identify PBC on the spin chain 
with PBC on the Ising model for $N$ even but anti-periodic boundary conditions on the Ising model for $N$ odd. 
  We have 
calculated the ground state energies in all 4 cases and the lowest 4 excited state energies for OBC and both parities of $N$; see Fig.~\ref{fig:energy}.  
 Note that, in stark contrast to the singlet sector,  the singlet-triplet gaps in Fig.~\ref{fig:energy}c and \ref{fig:energy}d go to a non-zero values at $1/N\to 0$.
This data on singlet energies determines  ten $x_n$ parameters. The nine parameters extracted from OBC all agree to within $5\%$ with the conformal field theory 
(CFT) predictions 
for the Ising model (see Table I in [\onlinecite{supmat}]). 
The agreement is not as good for PBC because the sizes accessible to DMRG are much
smaller. We plot the excited states energies in the upper panels of  Fig.~\ref{fig:energy}c and ~\ref{fig:energy}d.  The expected conformal tower structure of excited states is clearly revealed.\cite{supmat} Note that the extraction of the central charge from the entanglement entropy for PBC and OBC using the Calabrese-Cardy formula\cite{CalabreseCardy} is tricky because of the 
presence of strong oscillations,\cite{supmat} but the results are also consistent with $c=1/2$.

\begin{figure}[t!]
\includegraphics[width=0.4\textwidth]{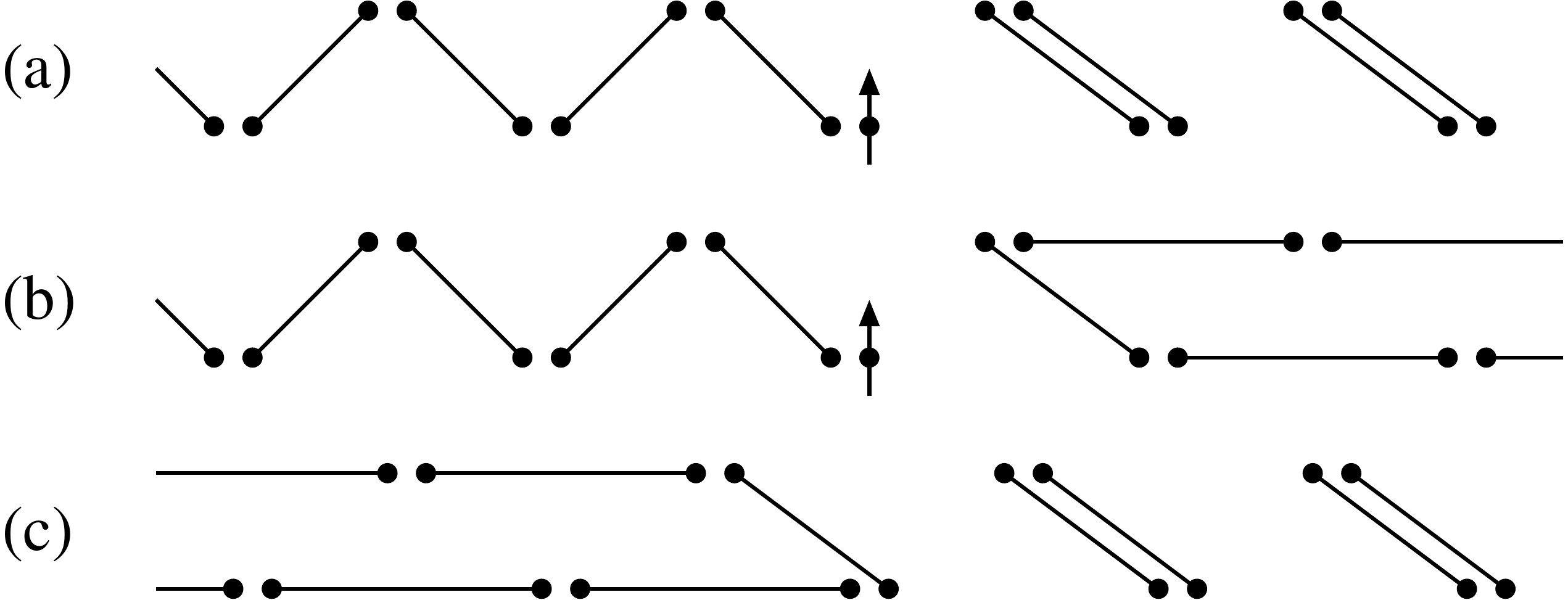}
\caption{Sketch of  domain walls between (a) the Haldane and Dimerized phases, (b) the Haldane and NNN Haldane phases, and (c) the NNN Haldane and Dimerized phases. A spin-1/2 appears at the domain wall in the first two cases,
but not in the third one.}
\label{fig:soliton}
\end{figure}

As stated above, the end point of the WZW SU(2)$_2$ is characterized by the absence of logarithmic 
corrections. So this is the only point along the line where the critical exponents can be accurately extracted 
from finite sizes. For the $SU(2)_2$ model,  CFT predicts $D(j,N)\propto 1/[(N/\pi) \sin (\pi x/N)]^{3/8}$. 
The `apparent' exponent decreases from $\simeq 0.43$ for $J_2=0$
until it reaches $3/8$ at $J_2\simeq0.12$ (see Fig.~\ref{fig:dimerization}(c)). As a confirmation, we have also extracted the
conformal towers at that point, for OBC with $N$ even or odd, and they fit well to the WZW SU(2)$_2$ prediction.\cite{supmat} 
Along the critical line, the central charge remains equal to $3/2$, including at the end point. The fact that it only drops around $J_2=0.2$
is presumably a finite size effect: since the gap is due to a marginal operator, it increases exponentially slowly above $J_2=0.12$ and
cannot be seen on small systems. 

To summarize, the spontaneous dimerization transition in spin-1 chains can be WZW SU(2)$_2$, Ising, or first-order depending on the parameters.
What is the rationale behind this unexpectedly rich situation? The first observation is that the WZW SU(2)$_2$ model has a marginal operator,
and in such a situation, the transition can be expected to turn first order if its coupling constant changes sign.
Regarding the alternative between Ising and WZW SU(2)$_2$, if the transition is continuous, we would like to suggest that it is intimately
related to the nature of the domain walls between the phases (see Fig.~\ref{fig:soliton}). 
A domain wall between Haldane and Dimerized phases necessarily carries spin-1/2 because the Haldane phase is topological
and has edge states, leading to a transition with magnetic excitations 
(WZW SU(2)$_2$ if it is continuous), whereas a domain wall between NNN-Haldane and Dimerized phases does not because the NNN-Haldane
phase is topologically trivial with no edge states, leading to an Ising transition
in the singlet sector with gapped magnetic excitations. These observations are consistent with the field theory approach. At a domain wall between 
Haldane and Dimerized phases $\theta (x)$ rotates by $\pm \sqrt{\pi}/2$, corresponding to $S^z=\pm 1/2$ excitations, whereas at a domain wall 
between NNN Haldane and Dimerized phases $\theta$ does not rotate.

The alternative between Ising and SU(2)$_2$ universality classes has been first pointed out by
Nersesyan and Tsvelik in the related context of spin-1/2 ladders with four-spin interactions using a Majorana fermion representation of the field theory.\cite{nersesyan,shelton}
Calculations on specific models with ring-exchange or frustrated leg coupling have supported this prediction.\cite{brehmer,nunner,muller,hijii_nomura,hijii_qin,schmidt,lauchli,gritsev,lecheminant_totsuka,fath,wang_2003,lavarelo,hijii_sakai} In that respect, the main difference with 
our model is that, in the model of Nersesyan and Tsvelik, one goes from Ising to SU(2)$_2$ through a trivial point of decoupled chains\cite{wang_2003,hijii_sakai} and central charge $c=2$, 
with no indication of an end-point of the SU(2)$_2$ line followed by a first-order transition, a generic feature of our approach due to the presence of
a marginal operator. 

Coming back to the role of edge states at the transition, the result summarized in Fig. \ref{fig:soliton} can easily be extended to ladders to explain the 
fundamental difference between Ising and SU(2)$_2$ universality classes:  spontaneous dimerization transitions between phases which are both topologically trivial (rung singlet and columnar dimer) or non trivial (Haldane and staggered dimer) can be expected to be generically Ising because edge states are absent or compensate each other, while spontaneous dimerization transitions 
between a topological and a non-topological phase (staggered dimer and rung singlet or Haldane and columnar dimer) 
must include magnetic excitations because of the edge states and can be expected to be 
generically SU(2)$_2$, or possibly first-order with spin-1/2 solitons. Similar ideas might be extended to transitions between valence-bond solids and dimerized phases in other contexts, possibly in higher dimension.

{\it Acknowledgments:} FM thanks A. L\"auchli for very insightful comments on the possible nature of the phase transitions, and C. Bazin, P. Lecheminant and A. Nevidomskyy for useful discussions. This work has been supported by the Swiss National Science Foundation and by NSERC Discovery Grant 36318-2009 and CIFAR (IA). 

\bibliographystyle{apsrev4-1}
\bibliography{bibliography}

\appendix

\section{Field Theory Approach}
We normalise the free boson Hamitlonian density:
\be {\cal H}={1\over 2}[(\partial_x \phi )^2+(\partial_x\theta )^2].\ee
The staggered components of the spin operators are represented in the conformal embedding by
\be \hbox{tr}g\vec \sigma \propto i\sigma (\sin \sqrt{\pi}\phi ,\cos \sqrt{\pi} \phi ,\cos \sqrt{\pi}\theta ).\label{gs}\ee
We see from this equation, and Eq. (3) of the paper, that $g$ has scaling dimension $1/8+1/4=3/8$, $(\hbox{tr}g)^2$ has dimension $1$ 
and $\vec J_L\cdot \vec J_R$ has dimension $2$, all correct values for the $SU(2)_2$ WZW model. The total central charge is $c=1+1/2=3/2$, 
also the correct value. Eq. (\ref{gs}) is also consistent with $\langle \hbox {tr}g\vec \sigma \rangle =0$ in all 3 phases, 
as must be the case since spin rotation symmetry is unbroken 
in all 3 phases. We also see from Eq. (\ref{gs}) that a spin rotation around the $z$ axis, $S^\pm_j\to e^{\pm i\alpha }S^\pm_j$, 
corresponds to $\phi \to \phi +\alpha /\sqrt{\pi}$, the $U(1)$ symmetry of the boson model. Thus all excitations of non-zero 
$S^z$ are in the boson sector. Since all  bosonic excitations are gapped on the Ising critical line, it follows from $SU(2)$ 
symmetry that all gapless excitations must have zero total spin on the Ising line. 

We note that $\theta$ and $\phi$ are not simply periodic bosons but rather 
$(\theta ,\sigma )$ should be identified with $(\theta +\sqrt{\pi},-\sigma )$ and $(\phi ,\sigma )$ should be identified with $(\phi +\sqrt{\pi},-\sigma )$. 
Therefore, for $\lambda_1<0$, there are only 2 inequivalent ground states, not 4, corresponding to the sign of $\langle \sigma \sin \sqrt{\pi}\theta \rangle$. 
In the Haldane phase where $\langle \sigma\rangle =0$, there is only 1 ground states with $\theta$ pinned at $0$ or $\sqrt{\pi}$ being equivalent. 
Likewise, in the NNN Haldane phase where $\langle \sigma\rangle =0$,  $\theta$ being pinned at $\pm \sqrt{\pi}/2$ are equivalent. 

The assumption that open boundary conditions in the Haldane phase impose a boundary condition $\theta (0)=\pm \sqrt{\pi}/2$ on the field theory may need 
further justification. This assumption can be justified close to the $c=3/2$ critical line by observing that the effective boundary magnetic field 
is $O(1)$ whereas the Haldane gap is very small. 

\section{Finite Size Spectrum}
There are 3 conformal towers that can occur in the finite size spectrum (FSS) of the Ising model, 
labeled by the corresponding primary fields, $I$, $\epsilon$ and 
$\sigma$, with dimensions $0$, $1/2$ and $1/16$ respectively. The finite size spectrum of the Ising model with 
the four different boundary conditions discussed in this paper were all worked out by Cardy.\cite{cardy86} With 
PBC and anti-periodic boundary conditions, direct products of conformal towers in left and right-moving sectors occur:
$(I,I)+(\epsilon ,\epsilon )+(\sigma ,\sigma )$ and $(I,\epsilon )+(\epsilon ,I)+(\sigma ,\sigma )$ respectively. 
The corresponding energies and momentum are:
\bea E&=&\epsilon_0N+{2\pi v\over N}\left[-{1\over 24}+x_R+x_L\right]\nonumber \\
P&=&{2\pi \over N}[x_R-x_L]
\eea
where $\epsilon_0$ is a non-universal constant and 
$x_R$ and $x_L$ are dimensions of chiral operators: dimensions of primary operators plus non-negative integers. The PBC ground state has $x_R=x_L=0$ and the 
anti-periodic boundary conditions ground state has $x_R=x_L=1/16$.  
For $\uparrow ,\uparrow$ boundary conditions only 
the conformal tower $I$ occurs and for $\uparrow ,\downarrow$ boundary conditions, only $\epsilon$ occurs. The corresponding 
finite size spectrum is:
\be E=\epsilon_0N+\epsilon_1+{\pi v\over N}\left[-{1\over 48}+x\right] \ee
where $\epsilon_1$ is another non-universal constant and $x$ is a dimension. 
The characters of these conformal towers, determining the multiplicities of excited states, 
 were calculated in [\onlinecite{rocha-caridi}]. The CFT predictions for the first few 
states in the spectrum, for all 4 boundary conditions are given in Table I and compared to our DMRG results. 

\begin{table}
\begin{tabular}{l||r|r}
 &  &DMRG\\
Energy level& CFT&$J_2=0.7$\\
 & Ising &$J_3=0.058$\\
\hline \hline
OBC, Even, ground state&-1/48& -1/48\\
\hline 
OBC, Even, 1st excited state&2& 1.99\\
\hline
OBC, Even, 2nd excited state&3& 2.90\\ 
\hline
OBC, Even, 3rd excited state&4& 3.82\\
\hline 
OBC, Even, 4th excited state&4& 3.87\\
\hline \hline 
OBC, Odd, ground state&23/48 $\simeq$  0.479& 0.477\\
\hline 
OBC, Odd, 1st excited state&1& 1.00 \\
\hline
OBC, Odd, 2nd excited state&2& 1.98 \\
\hline 
OBC, Odd, 3rd excited state&3& 2.98\\
\hline
OBC, Odd, 4th excited state&4& 3.97\\
\hline \hline 
PBC, Even, ground state&-1/12 $\simeq$ -0.0833& -0.094\\
\hline \hline
PBC, Odd, ground state&1/6 $\simeq$ 0.167 & 0.196
\end{tabular}
\caption{Energy levels on Ising line. Ground state refers to the $1/N$ term in the ground state energy.  For excited states, the gap above the ground state is given. 
Results are in units of $\pi v/N$. Note the degeneracy of the 3rd and 4th excited state, for OBC, N even, which occurs 
in the Ising conformal tower\cite{rocha-caridi} and is well-reproduced by our DMRG results.} 
\end{table}

\begin{table} 
\begin{tabular}{l||r|r}
 &  &DMRG\\
Energy level& CFT&$J_2=0.12$\\
 & SU(2)$_2$ &$J_3=0.087$\\
\hline \hline
OBC, Even, ground state $S_z^\mathrm{tot}=0$ &-1/16& -1/16\\
\hline 
OBC, Even, ground state $S_z^\mathrm{tot}=1$  &1& 1.027\\
\hline
OBC, Even, ground state $S_z^\mathrm{tot}=2$ &2& 2.052\\ 
\hline
OBC, Even, ground state $S_z^\mathrm{tot}=3$ &5& 5.14\\
\hline 
OBC, Even, ground state $S_z^\mathrm{tot}=4$ &8& 8.29\\
\hline 
OBC, Even, ground state $S_z^\mathrm{tot}=5$ &13& 13.50\\
\hline \hline 
OBC, Odd, ground state $S_z^\mathrm{tot}=1$  &7/16 & \\
&$\simeq$ 0.4375 & 0.443\\
\hline 
OBC, Odd, 1st exited state $S_z^\mathrm{tot}=1$  &1& 1.052\\
\hline
OBC, Odd, ground state $S_z^\mathrm{tot}=2$ &2& 2.052\\ 
\hline
OBC, Odd, ground state $S_z^\mathrm{tot}=3$ &4& 4.12\\
\hline 
OBC, Odd, ground state $S_z^\mathrm{tot}=4$ &8& 8.26\\
\hline 
OBC, Odd, ground state $S_z^\mathrm{tot}=5$ &12& 12.52
\end{tabular}
\caption{Energy levels at $SU(2)_2$ critical point. 
Ground state $S_z^\mathrm{tot}=0$ and  odd $S_z^\mathrm{tot}=1$ refers to the $1/N$ term in the ground state energy.  For the rest, the gap above the ground state is given. 
Results are in units of $\pi v/N$.} 
\end{table}

\begin{figure}[h!]
\includegraphics[width=0.47\textwidth]{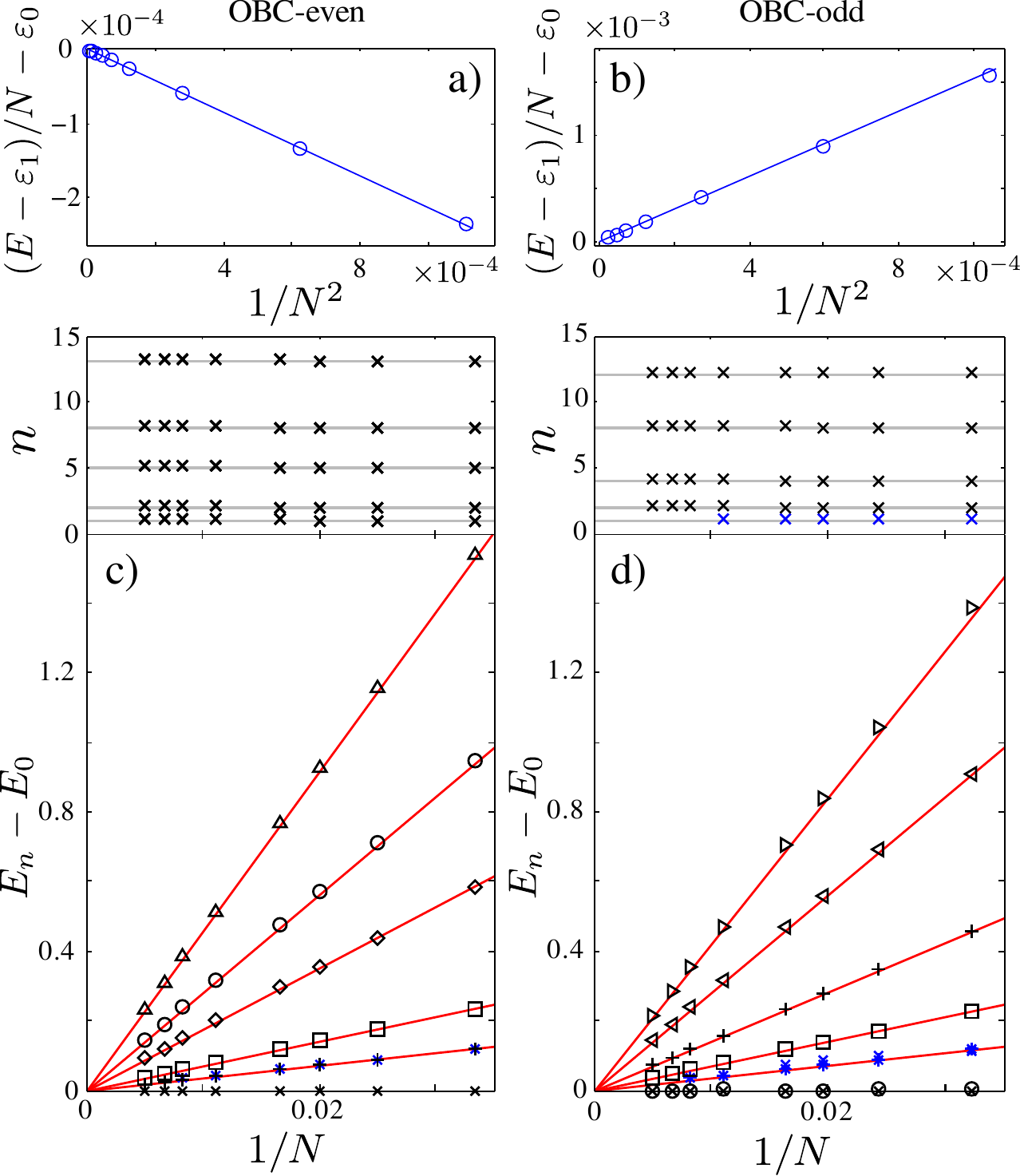}
\caption{(Color online) Ground state and excitation energy at $J_2=0.12$ and $J_3=0.087$, on the critical line between the Haldane and the Dimerized phases. Upper panels: Linear scaling of the ground state energy per site with $1/N^2$ after subtracting $\varepsilon_0$ and $\varepsilon_1$ in open chains with a) even and b) odd numbers of sites $N$. c) and d) Energy gap between the ground state and the lowest energies in different sectors of $S_z^\mathrm{tot}=0,1,...,5$ (black symbols) as a function of $1/N$ for even and odd numbers of sites. The multiplicity of the ground state and of the first excited states has
been obtained by calculating excited states in the sectors $S_z^\mathrm{tot}=0$ (blue crosses) and $S_z^\mathrm{tot}=1$ (blue pluses). Insets: Conformal towers for even and odd $N$. Black and blue symbols are DMRG data for the ground states in different sectors of $S_z^\mathrm{tot}$ and for the first excited state in the sector $S_z^\mathrm{tot}=1$}
\label{fig:ct_wzw}
\end{figure}

For the $SU(2)_2$ WZW model there are 3 conformal towers labeled by the spin of the lowest energy states, 
$j=0$, $1/2$ and $1$. Finite size spectra with conformally invariant boundary boundary conditions 
at both ends of the system can be determined from the corresponding boundary states, which are 
labeled by the primary operators.\cite{cardy89} OBC with $N$ even in our model  corresponds to the $|0\rangle$ 
boundary state at both ends of the system and the corresponding conformal tower in the FSS is $j=0$. Going to an 
odd number of sites is formally analogous to the infrared fixed point spectrum of a
 spin-1 Kondo model and the corresponding boundary state changes 
to $|1\rangle$ at one end of the system.\cite{affleckludwig91} The resulting FSS contains the $j=1$ conformal tower. 
Thus the ground state energies of an open chain with an even or odd number of sites are:
\begin{equation}
E_\mathrm{even}=\varepsilon_0 N+\varepsilon_1-\frac{\pi v}{16N},\ 
E_\mathrm{odd}=\varepsilon_0 N+\varepsilon_1+\frac{7\pi v}{16N}.
\end{equation}
In order to build the conformal tower at the end point $J_2=0.12$ and $J_3=0.087$, we calculate the gap between the ground state energy and the lowest energies in different sectors of $S_z^\mathrm{tot}$. The gap scales linearly with $1/N$ and the slope gives access to the velocity. 
By calculating excited states in the sectors $S_z^\mathrm{tot}=0,1$, we could determine the multiplicity of two lowest energy levels. In a chain with an
even number of sites, the ground state is a singlet and the first excited state is a triplet, while for a chain with an odd number of sites, the ground state is a triplet and the first excited state is degenerate and consists of one triplet and one singlet, in complete agreement with CFT predictions.
The DMRG data on the scaling is presented in Fig.\ref{fig:ct_wzw} a) and b) and summarised in Table II.

\begin{figure}[h!]
\includegraphics[width=0.49\textwidth]{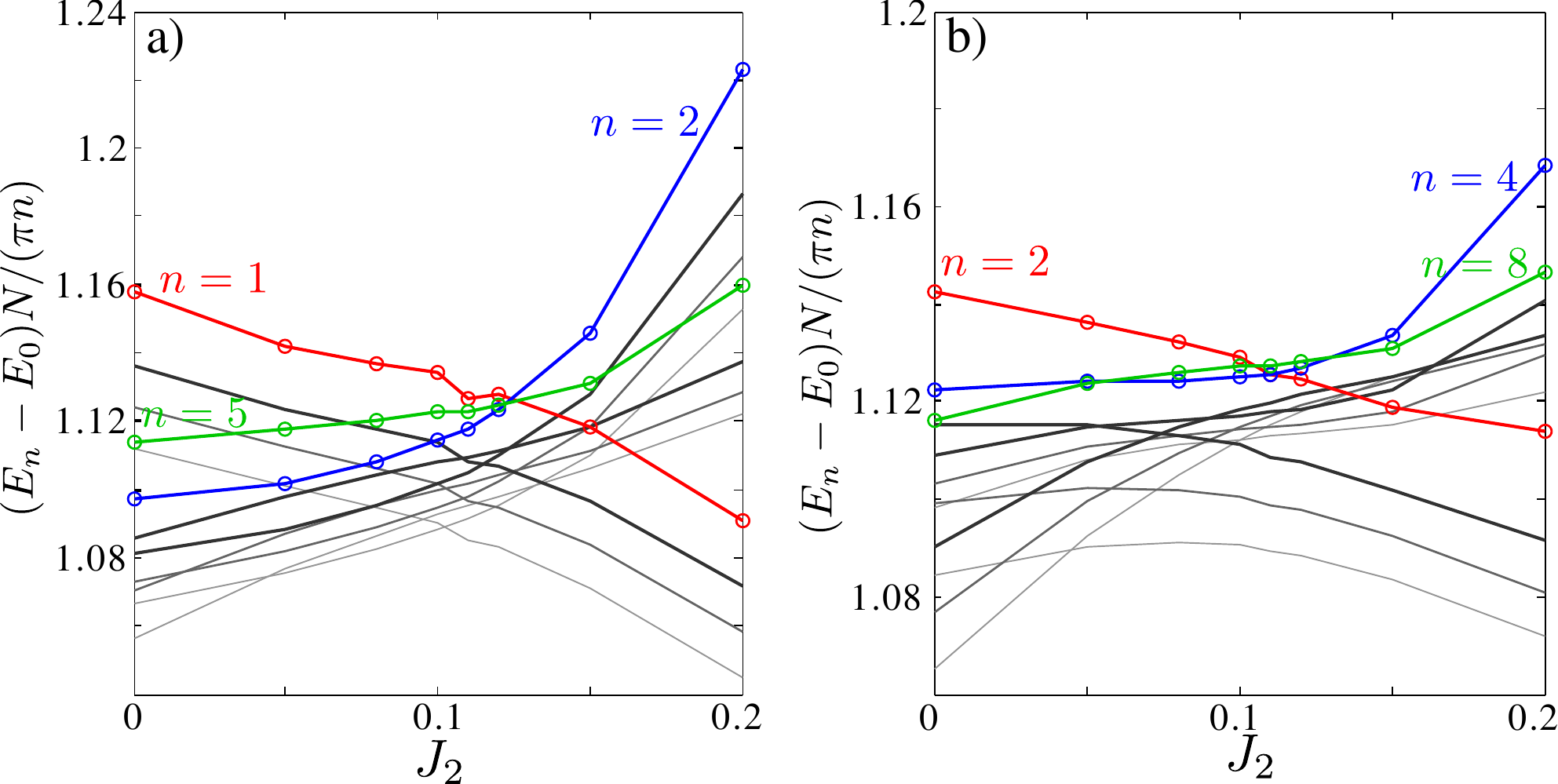}
\caption{(Color online) Velocity along the critical line between the Haldane and the dimerized phases extracted from the gap between $n$'s energy level and a groundstate. Red, blue and green lines show results for a) $N=50$  and b) $N=51$. Similar results for a) $N=30,24,20$ and b) $N=31,25,21$ are shown in gray  (from dark thick to light thin).}
\label{fig:velocity}
\end{figure}

We checked that the conformal tower is destroyed by moving along the critical line away from the end point. To demonstrate this, we have plotted the velocities extracted from three different excitation levels $n$ (fig.\ref{fig:velocity}). At the end point, all velocities are expected be the same, implying that the conformal tower is restored. This occurs around $J_2=0.12$, in agreement with the value determined from the critical exponent (see main text).

\section{Central charge from entanglement entropy at the Ising transition}

\begin{figure}[t!]
\includegraphics[width=0.45\textwidth]{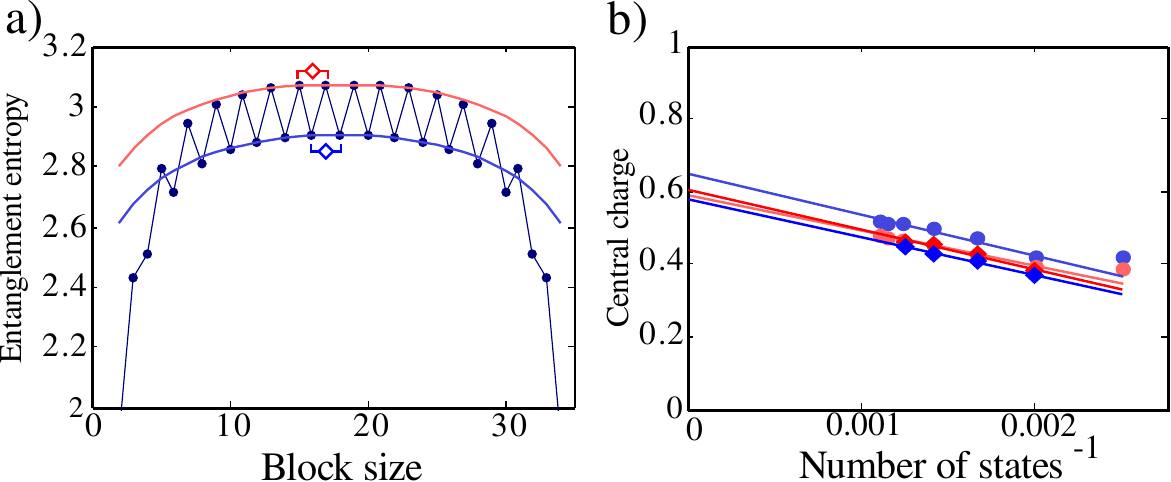}
\caption{(Color online) Extraction of the central charge for periodic boundary conditions and $J_2=0.7$, $J_3=0.058$. a) Example of entanglement entropy as a function of block size $n$ for $N=36$ sites and 800 states kept in DMRG. Light red and light blue lines are fits to the Calabrese-Cardy formula of Eq.~\ref{eq:calabrese_cardy}. Red and blue diamonds schematically show how the formula (\ref{eq:cc}) can be applied. b) Scaling of the central charge extracted in different ways with the number of states kept in the DMRG calculation. The lines are linear fits to the data-points (circles for the Calabrese-Cardy fit and diamonds for central charge calculated in the middle of the chain).}
\label{fig:cc_pbc}
\end{figure}

For a periodic chain with $N$ sites, the entanglement entropy of a subsystem of size $n$ is defined by $S_N(n)=-\mathrm{Tr}\rho_n\ln\rho_n$, where $\rho_n$ is the reduced density matrix. According to conformal field theory, the entanglement entropy in periodic systems depends on the size of the block according to:
\begin{equation}
S_N(n)=\frac{c}{3}\ln\left[\frac{N}{\pi}\sin\left(\frac{\pi n}{N}\right)\right]+s_1
\label{eq:calabrese_cardy}
\end{equation}
Let us first focus on the Ising transition between the NNN-Haldane and the dimerized phases. In the vicinity of this phase transition the convergence of the entanglement entropy in DMRG algorithm is very slow. This results in big oscillations that appear on top of the curve given by Eq.~\ref{eq:calabrese_cardy}. In principle these oscillations can be removed by increasing the number of sweeps and the number of states kept in DMRG. We went up to 16 sweeps keeping up to 900 states in two-site DMRG. With these parameters, oscillations disappear only for chains smaller than 30 sites. For larger systems, we have extracted the central charge for lower and upper curves of the entanglement entropy separately, as shown in Fig.~\ref{fig:cc_pbc}a). Note that the finite-size corrections to Eq.\ref{eq:calabrese_cardy} are minimal when the block sizev$n$ is as far as possible from the
extreme values $1$ and $N$\cite{franca}. Therefore we discard a few points close to the edges while fitting.
Alternatively, one can estimate the finite-size central charge by calculating it in the middle of the curve with only two points (see sketches with diamonds in Fig.~\ref{fig:cc_pbc}a)). Using Eq.~\ref{eq:calabrese_cardy}  
leads to the estimates:
\begin{equation}
c_{k}=\frac{3\left[S_N(\frac{N}{2}-(k+2))-S_N(\frac{N}{2}-k)\right]}{\ln\left[\cos(\frac{(k+2)\pi}{N})/\cos(\frac{k\pi}{N})\right]},
\label{eq:cc}
\end{equation}
where $k=0,1$ for upper and lower curves.

\begin{figure}[t!]
\includegraphics[width=0.3\textwidth]{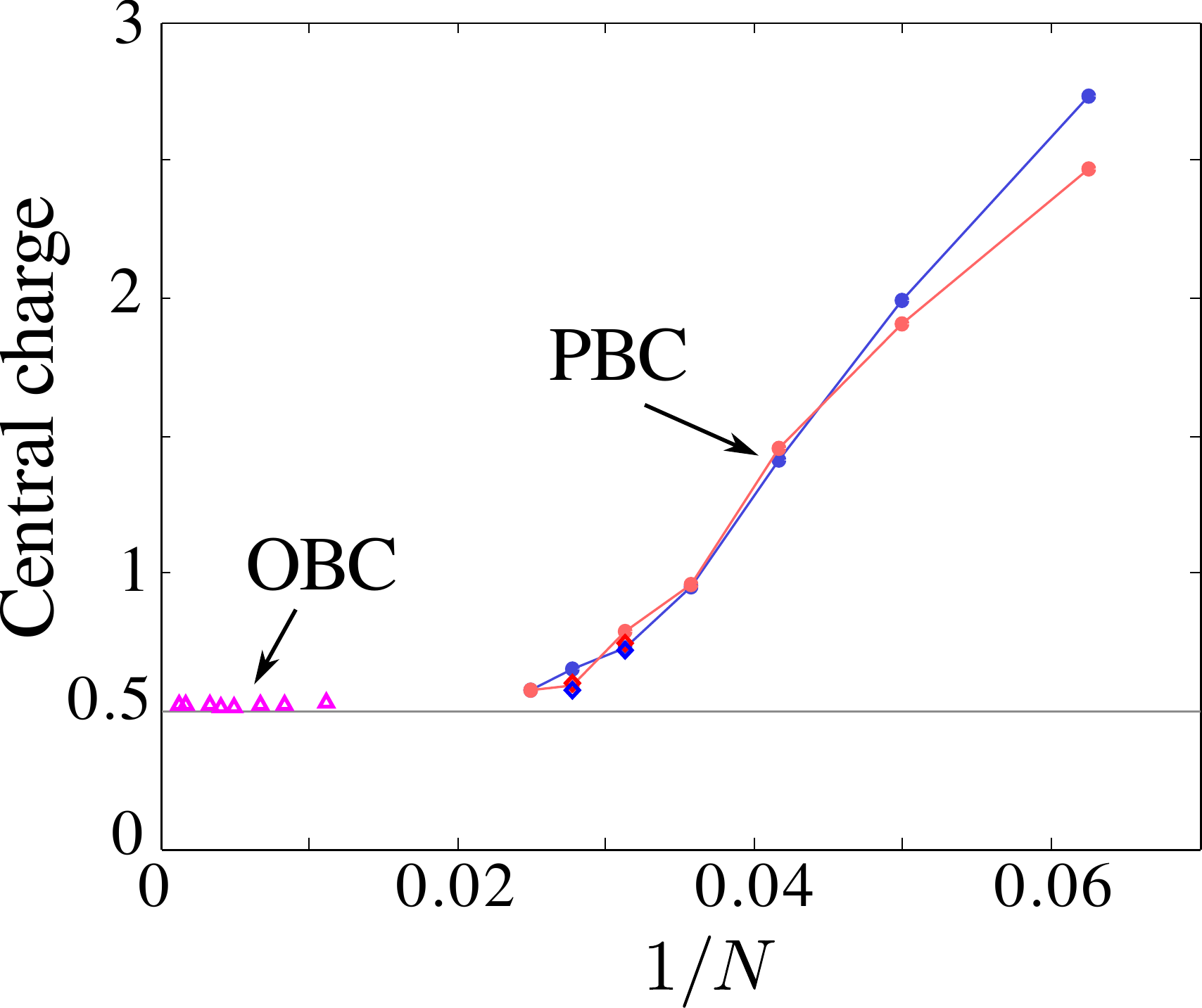}
\caption{(Color online)  Central charge for the Ising transition as a function of $1/N$. The light blue and red circles have been obtained with the fits of the upper and lower curves of the entanglement entropy with Calabrese-Cardy formula. The red and blue diamonds stand for the central charge extracted in the middle of each curve. All results are extrapolated with the inverse number of sweeps. Magenta triangles stand for the central charge extracted from the entanglement entropy in open chains.}
\label{fig:cc_summary}
\end{figure}

For each system size, we then extrapolate the extracted values of the central charges with the number of states kept in DMRG algorithm. The extrapolated values of the central charge as a function of system size $N$ are shown in Fig.~\ref{fig:cc_summary}. They are consistent with $c=1/2$ in the thermodynamic limit.
 
\begin{figure}[t!]
\includegraphics[width=0.45\textwidth]{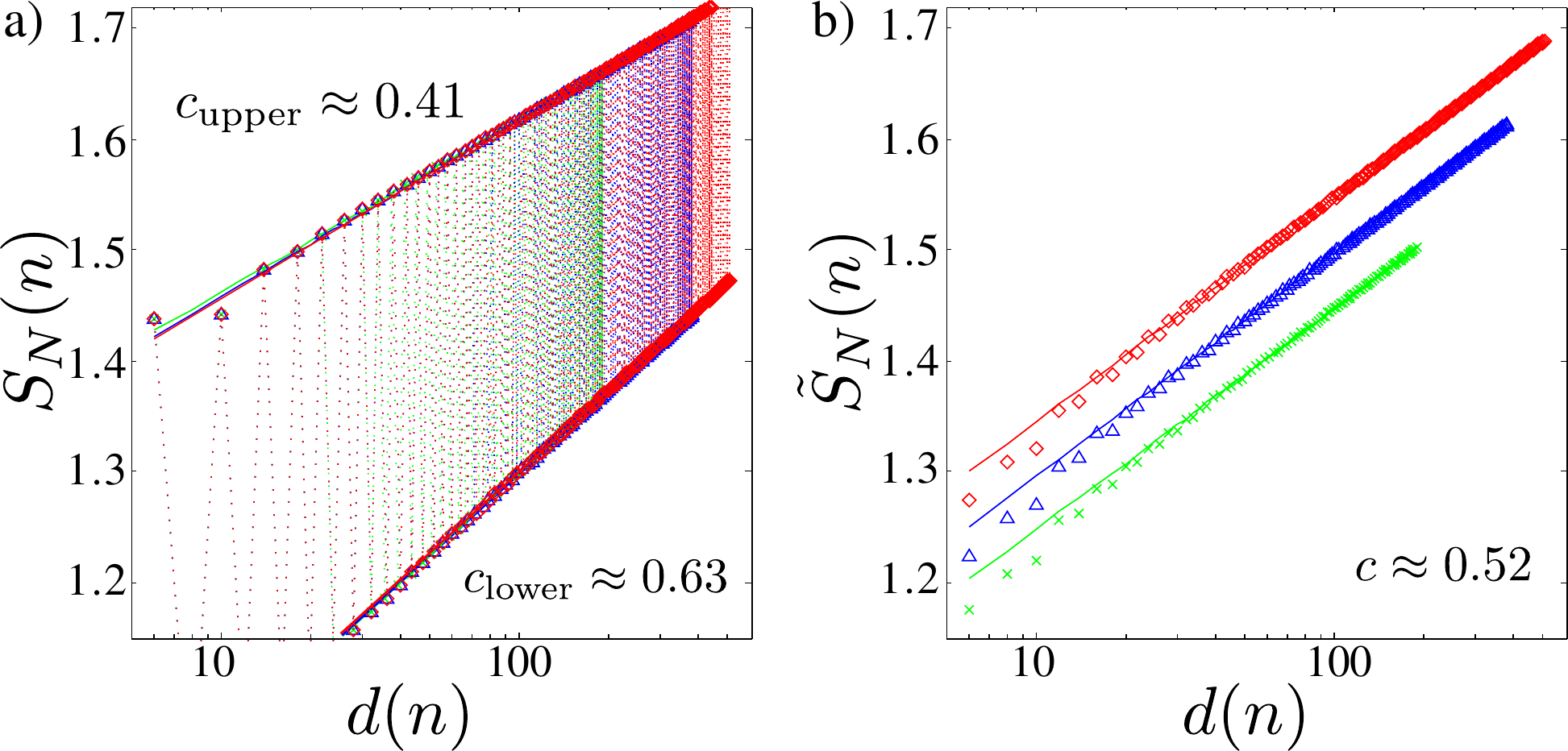}
\caption{(Color online) Entanglement entropy as a function of the conformal distance for $N=300$ (green), $600$ (blue) and $800$ (red) sites at $J_2=0.7$ and $J_3=0.058$. a) The solid lines are fits of the upper and lower curves to Eq.~\ref{eq:calabrese_cardy_obc}. The slopes of the fits give upper and lower limits for the central charge. b) Entanglement entropy after removing the Friedel oscillations with weight $\zeta\approx 2/9$. The data for $N=300$ and $600$ are shifted downward by 0.1 and 0.05 for clarity. }
\label{fig:cc_obc}
\end{figure}

\begin{figure}[t!]
\includegraphics[width=0.45\textwidth]{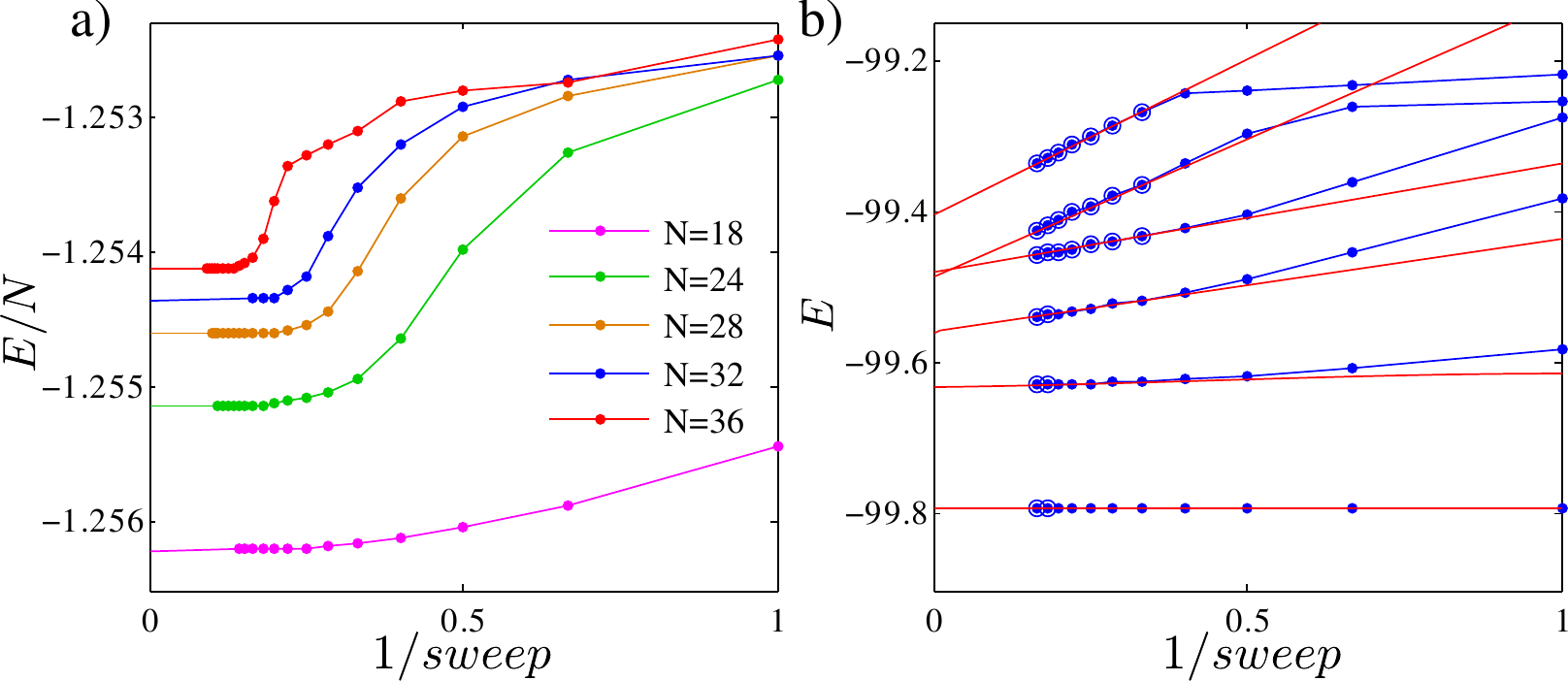}
\caption{(Color online) Extrapolation of the DMRG results towards infinite number of sweeps. a) Ground state energies for periodic chains with different numbers of sites. The continuation of the line is a fit linear in $1/sweep$ of the last few points. b) Ground-state energy and energy of the lowest excited states as a function of the inverse number of DMRG sweeps. Dots are DMRG results while red lines are linear fits of the last few points for each curve marked with large circles.}
\label{fig:convergence}
\end{figure}

It is well established that DMRG algorithm has better performances for open systems, and much bigger system sizes can be reached then. In systems with open boundary conditions, the entanglement entropy scales with the block size according to:
\begin{equation}
S_N(n)=\frac{c}{6}\ln\left[\frac{2N}{\pi}\sin\left(\frac{\pi n}{N}\right)\right]+s_1+\log g
\label{eq:calabrese_cardy_obc}
\end{equation}
Since we are dealing here with much larger system sizes it is useful to present results in a logarithmic scale by introducing the conformal distance $d$:
\begin{equation}
d=\frac{2N}{\pi}\sin\left(\frac{\pi n}{N}\right)
\label{eq:conformal_distance}
\end{equation}
As in the case of periodic boundary conditions, big oscillations appear on top of the prediction of Eq.~\ref{eq:calabrese_cardy_obc}. However, in open systems, the oscillations are caused by Friedel oscillations and cannot be removed by increasing the number of sweeps or the number of states. Separate fits of the upper 
and lower curves of the entanglement entropy leads to rough estimates of the central charge: $c_\mathrm{lower}\approx0.41$ and $c_\mathrm{upper}\approx0.63$ (see Fig.~\ref{fig:cc_obc}a)).

In order to remove the oscillations, following Ref.~[\onlinecite{capponi}], we have subtracted the spin-spin correlation on the corresponding link from the entanglement entropy with some weight $\zeta$. Then the reduced entanglement entropy as a function of the conformal distance takes the form:
\begin{equation}
\tilde{S}_N(n)=\frac{c}{6}\ln d(n)+\zeta \langle {\bf S}_n{\bf S}_{n+1} \rangle+s_1+\log g
\label{eq:calabrese_cardy_obc_corrected}
\end{equation}
The results of the numerical calculation of the central charge from the entanglement entropy for both OBC and PBC are summarized in 
Fig~\ref{fig:cc_summary}. These results are consistent with $c=1/2$.

\section{Convergence of energies in DMRG}

Close to the critical point, the DMRG algorithm converges very slowly, especially for periodic boundary conditions.
In Fig.~\ref{fig:convergence}a), we have plotted the ground-state energy of periodic chains as a function of the inverse number of sweeps. Note that we plot
measurements after each passage through the system, whereas a sweep corresponds to going back and forth. So the variable sweep takes half-integer as 
well as integer values. The almost flat part of the curves for large number of sweeps indicates that convergence was reached. For each curve, we have used the slope of the last few points to extrapolate the results for infinite number of sweeps. We do up to 16 sweeps and keep up to 900 states. In the first 6-7 sweeps the number of kept states increases from 100 to approximately $90\%$ of the maximal value, in the following sweeps we jiggle the wave-function by decreasing and increasing the number of states until the convergence is reached.

The lack of convergence is also a problem for higher excited states even with open boundary conditions as shown in Fig.~\ref{fig:convergence}b). To estimate
the excitation energies, we have extrapolated the last few points of each curve to infinite number of sweeps with a linear fit in $1/sweep$. We do 7-9 sweeps and the number of kept states increase linearly from 100 to 900. Therefore finite-size scaling of the energy with the number of sweeps is equivalent to scaling with number of kept states.

\end{document}